\documentclass{article}
\usepackage[ansinew]{inputenc}
\usepackage{latexsym}
\newcommand{\be}{\begin{equation}}
\newcommand{\ee}{\end{equation}}
\newcommand{\bear}{\begin{eqnarray}}
\newcommand{\ear}{\end{eqnarray}}
\newcommand{\ba}{\begin{eqnarray*}}
\newcommand{\ea}{\end{eqnarray*}}

\newcommand{\dt}{\mbox{\boldmath$:$}}

\newcommand{\no}{\noindent}

\begin{document}

\title{Thermofield Quantum Electrodynamics in $1 + 1$ Dimensions at Finite Chemical Potential: A  Bosonization Approach}
\author{ R. L. P. G. Amaral$^a$, L. V. Belvedere$^a$ and K. D. Rothe$^{b}$\footnote{Email addresses, respectively: rubens@if.uff.br, belve@if.uff.br and K.Rothe@thphys.uni-heidelberg.de.} \\
\small{${}^a$ Instituto de F\'{\i}sica}\\
\small{Universidade Federal Fluminense}\\
\small{Av. Litor\^anea S/N, Boa Viagem,  CEP. 24210-340}\\
\small{Niter\'oi - RJ - Brasil}\\
\small{${}^b $ Institut f\"ur Theoretische Physik}\\
\small{Universit\"at Heidelberg}\\
\small{Philosophenweg 16, D-69120}\\
\small{ Heidelberg - Germany}\\
}

\date{November 24, 2010}

\maketitle
\begin{abstract}
The recent generalization of the Lowenstein-Swieca operator solution of Quantum Electrodynamics in 1+1 dimensions to finite temperature
in Thermofield Dynamics is further generalized to include a non-vanishing chemical potential. The operator solution to the Euler-Lagrange equations respecting the Kubo-Martin-Schwinger condition is constructed. Two forms of this condition and their associated solutions are discussed. The correlation functions of an arbitrary number of chiral densities are computed in the thermal $\theta$-vacuum.
 
\end{abstract}

\section{Introduction}

The bosonization of fermion fields is a very useful technique for solving models in 1+1 dimensions, and also provides a very instructive framework for studying non-perturbative aspects \cite{AAR}. 

In \cite{ABRI} we have considered the bosonic operator representation of massless free fermions at finite temperature (thermofield bosonization) within the formalism of Thermofield Dynamics \cite{TFD,U,Das,Mats1,O,Mats}. It was shown that the  well known two-dimensional Fermion-Boson correspondences at zero temperature also hold at finite temperature. Using this thermofield bosonization we solved the massless Thirring model at finite temperature \cite{ABRI}. In ref. \cite{ABRII} we then extended the thermofield bosonization approach of ref. \cite{ABRI} to the case of massive free fermions, and computed the corresponding $N$-point correlation functions of chiral densities as a perturbative series in the mass. By working in the interaction picture, the infinite series in the mass parameter of the fermionic formulation was compared to the corresponding perturbative series in the interaction parameter of the bosonized thermofield formulation. In this way, we established in thermofield dynamics the formal equivalence of the massive free fermion theory to the sine-Gordon dynamics at a particular value of the sine-Gordon parameter \cite{C}.

The intimate relationship between the thermofield dynamics formalism and the algebraic formulation due to Haag-Hugenholtz-Winnink (HHW) of statistical mechanics \cite{Haag} has been established in Ref. \cite{O}. In this reference the relevance of tilde objects to the modular conjugation appearing in the algebraic formulation of statistical mechanics in the HHW formalism based upon the Kubo-Martin-Schwinger (KMS) condition \cite{KMS} was clarified, and a revised version of the thermofield dynamics approach for fermions was presented. In Ref. \cite{ABRII} we have shown that this revised version was fundamental in order to establish the thermofield bosonization scheme for the free massless Fermi field.

Quantum Electrodynamics in 1+1 dimensions ($QED_2$) has been solved on operator level in a classical paper by Lowenstein and Swieca \cite{Lowenstein}.
In a recent paper \cite{ABRIII} we have extended the Lowenstein-Swieca operator solution  to the case of non-zero temperature using 
Thermofield Dynamics \cite{U}. In particular we have seen that despite the doubling of the Hilbert space required by this formalism, the infinity of ``theta vacua"
characteristic of $QED_2$ at zero temperature is not doubled. 

In the present paper we extend the solution obtained in \cite{ABRIII} to include
a non-vanishing chemical potential. The paper is organized as follows: 

In section 2 we begin by considering the free massless fermion field and compute in Thermofield Dynamics the expectation value of 2n-point functions
as the thermodynamic average with a  Boltzmann distribution. The results show an oscillatory behaviour in the chemical potential and obey the Kubo-Martin-Schwinger (KMS) \cite{KMS} condition. We also show what this implies on the bosonized level. Our results, obtained in Thermofield Dynamics, are found to be in agreement with those of ref. \cite{Min},
obtained in a functional approach. We further generalize these results to the case of independent chemical potentials
for each chiral component.

In section 3 we turn to $QED_2$,  making use of the results of section 2 and imposing the KMS condition in order to introduce the chemical potential. Our 
conclusion is in agreement with the starting point of refs. \cite{Kogut,Kao,Schaposnik} and the prescription given in ref. \cite{Actor}. We then construct the corresponding
operator solution of the Dirac and Maxwell equations. Besides the well known fact that the latter can be satisfied only in the weak sense, a short distance calculation implying a non-vanishing vacuum expectation value of the current shows that the Maxwell equation can only be satisfied  provided we introduce a constant background distribution in the Lagrangian,
as witnessed already in the work of refs. \cite{Kogut,Kao}.

In section 4 we finally compute in thermal $QED_2$ the correlation functions of products of chiral densities and compare the results with those of \cite{Schaposnik}, obtained at zero temperature and finite chemical potential
from the functional point view, when taking account of the topologically non-trivial gauge field configurations. Our results obtained on operator level
in Thermofield Dynamics display the relative simplicity of these calculations, and at the same time generalize the 
( implicitly)  $\theta=0$ vacuum calculations at zero temperature of \cite{Schaposnik} to arbitrary ``theta" vacua at finite temperature.

\section{Two-dimensional free massless Fermi theory at finite density in the Thermofield Dynamics approach}
\setcounter{equation}{0}

Within the Thermofield Dynamics approach \cite{TFD,U,Das,Mats1,O,Mats}, the formulation of a Quantum Field Theory at finite temperature in terms of operators requires doubling the number of field degrees of freedom. This is achieved by introducing  a fictitious ``tilde'' operator for each of the operators describing the system under consideration, and thus entails a doubling of the Hilbert space. The fictitious ``tilde" system is an copy of the original system under consideration with ``canonical" 
commutation relations of opposite signature.  To begin with, we consider the case of two-dimensional massless fermions at zero temperature and zero chemical potential, described by the doublets 
\footnote{Left and right moving fields are only distinguished by their arguments. Our conventions are the same as in 
\cite{ABRIII}.}
\[
\psi ^{(0)} (x)\,=\,\pmatrix{\psi^{(0)} (x^+) \cr \psi^{(0)} (x^-)}\,,\quad \widetilde\psi^{(0)}(x)\,=\,\pmatrix{\tilde{\psi}^{(0)} (x^+) \cr\tilde{ \psi}^{(0)} (x^-)}
\]
where $x^\pm = x^0 \pm x^1$, whose dynamics in the doubled Hilbert space is described by the total Lagrangian
\[
\hat{\cal L}_0 = {\cal L}_0-\tilde{\cal L}_0\,,
\]
with
\[
{\cal L}_{0} = \bar\psi^{(0)}(x)i\gamma^\nu\partial_\nu \psi^{(0)}(x)\,,\quad \tilde{\cal L}_0 =- \bar{\tilde{\psi}}{}^{(0)}(x)i\gamma^\nu\partial_\nu {\tilde{\psi}}^{(0)}(x)\,,
\]
and where the ``tilde'' conjugation operation is defined for a general operator ${ A}$ by $\widetilde{c A} = c^\ast \widetilde A$, with c a c-number. At finite inverse temperature $\beta$ and chemical potential $\mu$ the thermal averages of field operators $A$ are given by

\be\label{Boltzmann-average}
\langle A \rangle   = \frac{Tr  Ae^{-\beta(H_0 - \mu {\cal Q})}}{Tr e^{-\beta(H_0 - \mu {\cal Q})}},
\ee

\no with $H_0$  the Hamiltonian
\[
H_0 = \int_{-\infty}^\infty {dp^1}\vert p^1\vert [b^\dagger(p^1)b(p^1) + d^\dagger(p^1)d(p^1)]\,, 
\]
and $Q$ the charge operator
\be\label{charge-operator}
{\cal Q} = \int_{-\infty}^\infty {dp^1}[b^\dagger(p^1)b(p^1) - d^\dagger(p^1)d(p^1)] .
\ee

\no  Here $b(p^1)$ and $d(p^1)$  ($b^\dagger(p^1)$ and $d^\dagger(p^1)$) are the usual fermion and anti-fermion destruction (creation) operators appearing in the Fourier representation of the chiral components of the Fermion field ($p = \vert p^1 \vert = p^0 $)
\[
\psi^{(0)} (x^\pm )\,=\,\int_0^\infty dp \,\Big \{\,f_p (x^\pm)\, b (\mp p)\,+\,f^\ast_p (x^\pm)\, d^\dagger (\mp p)\,\Big \}\,,
\]
 where
\[
f_p (x^\pm)\,=\frac{1}{\sqrt{2\pi}}\,e^{\,-\,i\,p x^\pm}\,,
\]
with a corresponding expression for the tilde-fields obtained by the tilde conjugation rule. 

In order to compute the fermionic two-point function at finite temperature we shall use the revised thermofield dynamics approach due to Ojima \cite{O,ABRI,ABRII,ABRIII}  modified to include the chemical potential. In this approach the thermal average (\ref{Boltzmann-average}) of an observable  can be expressed as the expectation value of the corresponding operator $A$ in a thermal vacuum 
$\vert 0(\beta,\mu) \rangle$  obtained by the unitary transformation
\be\label{beta-vacuum}
\vert 0(\beta,\mu) \rangle = {\cal U}_F\,[\theta_F(\beta,\mu)] \vert 0,\tilde 0 \rangle\,,
\ee
in the doubled Hilbert space, where the unitary operator ${\cal U}_F [\theta_F(\beta ,\mu) ]$ is given by

\be\label{unitary}
{\cal U}_F [\theta_F(\beta ,\mu) ]\,=\,e^{ \, -\,i\,\int_{- \infty}^{+ \infty}\,d p^1\,\big [ \big ( \tilde b (p^1 ) b (p^1)\,
+\,b^\dagger (p^1) \tilde b^\dagger (p^1) \big ) \theta ( | p^1|  ; \beta,-\mu )\,
+\,\big (\tilde d (p^1 ) d (p^1)\,+\,d^\dagger (p^1) \tilde d^\dagger (p^1) \big ) \theta (|p^1| ;\beta,\mu ) \big ) \big ]} \,,
\ee
with the Bogoliubov parameters $\theta (\beta,\mu )$ implicitly defined by

\be\label{coseno}
\cos \theta ( p ; \beta,\mu )\,=\,\frac{1}{\sqrt{ 1\,+\,e^{-\,\beta (p+ 
\mu )}}}\,,
\ee
\be\label{seno}
\sin \theta (p ; \beta,\mu )\,=\,\frac{e^{\,-\,\frac{\beta}{2} (p+ \mu )}}{\sqrt{ 1\,+\,e^{-\,\beta (p+\mu )}}}\,,
\ee
with $\mu = \mu_F \in \Re$ and the corresponding Fermi-Dirac statistical weight  \cite{Das}
\[
\sin^2 \theta (p ; \beta,\mu ) \,=\,\frac{\displaystyle 1}{\displaystyle 1\,+\,e^{\,\beta ( p + \mu ) }}\,,
\]

\no The transformed annihilation operators are given by

\be\label{b}
b (p ; \beta , \mu )\,=\,{\cal U}^{-1}_F [\theta(\beta ,\mu ) ]\,b (p)\,{\cal U}_F [\theta (\beta ,\mu ) ]\,=\,
b (p) \cos {\theta (p ; \beta , -\mu )}\,+\,i\,\tilde b^\dagger (p) \sin {\theta (p ; \beta , -\mu )}\,,
\ee

\be\label{d}
d (p ; \beta, \mu )\,=\,{\cal U}^{-1}_F [\theta (\beta,\mu ) ]\,d (p)\,{\cal U}_F [\theta(\beta,\mu ) ]\,=\,
d (p) \cos {\theta (p ; \beta , \mu )}\,+\,i\,\tilde d^\dagger (p) \sin {\theta (p ; \beta , \mu )}\,.
\ee

\no The operators $\tilde b (p ; \beta , \mu )$ and $\tilde d (p ; \beta , \mu )$ are obtained from the  operators above by the tilde conjugation rule, i.e. $\widetilde{c b (p)} = c^\ast \tilde b (p)$
and $\widetilde{c d (p)} = c^\ast \tilde d (p)$. From here and (\ref{beta-vacuum})
the non-vanishing diagonal statistical ensemble averages are found to be
(To simplify the notation, $p$ and $k$ stand in the following for $p^1$ and $k^1$, respectively.)

\[
\langle 0_{_{F}} (\beta ,\mu ) \vert b (p ) b^\dagger (k ) \vert \ 0_{_{F}} (\beta ,\mu ) \rangle\,=\,
\frac{1}{1\,+\,e^{\,-\,\beta ( |p| - \mu)}}\,\delta (p - k)\,,
\]

\be\label{diagonal}
\langle 0_{_{F}} (\beta ,\mu ) \vert b^\dagger (p ) b (k ) \vert  0_{_{F}} (\beta ,\mu ) \rangle\,=\,\frac{e^{\,-\,\beta ( |p| - \mu )}}{1\,+\,e^{\,-\,\beta ( |p| - \mu)}}\,\delta (p - k)\,,
\ee

\[
\langle 0_{_{F}} (\beta ,\mu ) \vert d (p ) d^\dagger (k) \vert 0_{_{F}} (\beta ,\mu ) \rangle\,=\,\frac{1}{1\,+\,e^{\,-\,\beta ( |p| + \mu)}}\,\delta (p - k)\,,
\]

\[
\langle 0_{_{F}} (\beta ,\mu ) \vert d^\dagger (p ) d (k ) \vert  0_{_{F}} (\beta ,\mu ) \rangle\,=\,\frac{e^{\,-\,\beta ( |p| + \mu )}}{1\,+\,e^{\,-\,\beta ( |p| + \mu)}}\,\delta (p - k)\,,
\]

\no and the same for the corresponding diagonal  averages of the  ``tilde'' operators. For the off-diagonal non-vanishing statistical averages we obtain

\[
\langle 0_{_{F}} (\beta ,\mu ) \vert b (p ) \,i\,\tilde b (k ) \vert  0_{_{F}} (\beta ,\mu ) \rangle\,=\,
\frac{e^{\,-\,\frac{\beta}{2} ( |p| - \mu )}}{1\,+\,e^{\,-\,\beta ( |p| - \mu)}}\,\delta (p - k)\,,
\]

\be\label{off-diagonal}
\langle 0_{_{F}} (\beta ,\mu) \vert \,i\,\tilde b (p ) b (k ) \vert  0_{_{F}} (\beta ,\mu) \rangle \,= \,-\,
\frac{e^{\,-\,\frac{\beta}{2} ( |p| - \mu )}}{1\,+\,e^{\,-\,\beta ( |p| - \mu)}}\, \delta (p - k)\, ,
\ee

\[
\langle 0_{_{F}} (\beta ,\mu) \vert d (p ) \,i\,\tilde d (k ) \vert  0_{_{F}} (\beta  ,\mu) \rangle\,=\,\frac{e^{\,-\,\frac{\beta}{2} ( |p| + \mu )}}{1\,+\,e^{\,-\,\beta ( |p| + \mu)}} \,\delta (p - k) ,
\]

\[
\langle 0_{_{F}} (\beta ,\mu) \vert \,i\,\tilde d (p ) d (k ) \vert  0_{_{F}} (\beta ,\mu) \rangle = \,-\,\frac{e^{\,-\,\frac{\beta}{2} ( |p| + \mu )}}{1\,+\,e^{\,-\,\beta ( |p| + \mu)}} \,\delta (p - k)\, .
\]

Now, the Fermi thermofields are given by

$$
\psi^{(0)}(x^\pm ; \beta,\mu )\,=\,\int_0^\infty dp \,\Big \{\,f_p (x^\pm)\,\Big ( b (\mp p)\,\cos \theta ( p ; \beta,-\mu )\,+\,i\,\widetilde b^\dagger (\mp p)\,\sin \theta ( p ; \beta,-\mu )\,\Big )\,+
$$

\be\label{f1}
f^\ast_p (x^\pm)\,\Big ( d^\dagger (\mp p)\,\cos \theta ( p ; \beta,\mu )\,-\,i\,\widetilde d (\mp p)\,\sin \theta ( p ; \beta,\mu )\,\Big )\,\Big \}\,,
\ee

$$
\widetilde\psi^{(0)}(x^\pm ; \beta,\mu )\,=\,\int_0^\infty dp\, \Big \{\,f^\ast_p (x^\pm)\,\Big ( \widetilde b (\mp p)\,\cos \theta ( p ; \beta;-\mu )\,-\,i\, b^\dagger (\mp p)\,\sin \theta( p ; \beta,-\mu )\,\Big )\,+
$$

\be\label{f2}
f_p (x^\pm)\,\Big ( \widetilde d^\dagger (\mp p)\,\cos \theta ( p ; \beta,\mu )\,+\,i\, d (\mp p)\,\sin \theta ( p ; \beta,\mu )\,\Big )\,\Big \}\,.
\ee

With (\ref{diagonal}) and (\ref{off-diagonal}) we can now calculate the two-point functions  of the fermionic doublet. As we shall see, the fact that the Fermi fields under consideration are massless will enable one to factorize the chemical potential dependence of the two-point functions into an exponential pre-factor. The diagonal two-point function is given by

\[
\langle 0_F (\beta,\mu ) \vert \psi^{(0)} (x^\pm) \psi^{(0)\dagger} (y^\pm) \vert 0_F (\beta,\mu ) \rangle\,=\,\langle 0 , \widetilde 0 \vert \psi^{(0)} (x^\pm ; \beta,\mu ) \psi^{(0)\dagger} ( y^\pm ; \beta,\mu ) \vert \widetilde 0 , 0 \rangle
\]

\be \label{diag}
=\frac{1}{2 \pi}\,\int_0^\infty\,dp\,\big \{ e^{\,-\,i\,p\,(x^\pm\,-\,y^\pm )}\, \cos^2 \theta (p; \beta,-\mu )\,+\,e^{\,+\,i\,p\,(x^\pm\,-\,y^\pm)}\, \sin^2 \theta (p; \beta,\mu )\,\big \}\,.
\ee

\no Using eqs. (\ref{coseno}, \ref{seno}) and performing a change of variables $p - \mu  \rightarrow p$, we obtain

\begin{eqnarray}\label{diag2}
\langle0(\beta,\mu)\vert  \psi^{(0)} (x^\pm) \psi^{(0)\dagger}(y^\pm) \vert 0(\beta,\mu)\rangle\,&=&\,\int_0^\infty\,\frac{dp}{2\pi}\, \left\{ e^{\,-\,i\,p\,(x^\pm - y^\pm)}\, \frac 1{1+e^{-\beta(p-\mu)}}\,+\,e^{\,\,i\,p\,(x^\pm-y^\pm) }\, \frac 1{1+e^{\beta(p+\mu)}}\, \right\}\,\nonumber\\
&=&\,\int_{-\infty}^\infty\,\frac{dp}{2\pi}\,\left\{ e^{\,-\,i\,p\,(x^\pm-y^\pm)}\, 
\frac 1{1+e^{-\beta(p-\mu)}}\,\right\}\,\nonumber\\
&=&\,e^{-i\mu ( x^\pm-y^\pm)}\,\,\int_{-\infty}^\infty\,\frac{dp}{2\pi}\,\left\{ e^{\,-\,i\,p\,(x^\pm-y^\pm)}\, \frac 1{1+e^{-\beta p}}\,\right\}\,,
\end{eqnarray}

\no The last integral in eq. (\ref{diag2}) corresponds to the diagonal thermal two-point function at zero chemical potential \cite{ABRI},

\be\label{psi-correlator}
\langle 0_F (\beta ) \vert \psi^{(0)} (x^\pm ) \psi^{(0)\dagger} ( y^\pm) \vert  0_F (\beta ) \rangle\,
=\,\frac{\displaystyle 1}{\displaystyle 2 i \beta\,\sinh \frac{\pi}{\beta}\,(x^\pm\,-\,y^\pm\,-\,i\,\epsilon )}\,.
\ee

\no In this way we conclude from (\ref{diag2}) and (\ref{psi-correlator}) that for $\mu\neq 0$,
\be\label{d2pf1}
\langle 0_F (\beta,\mu ) \vert \psi^{(0)} (x^\pm) \psi^{(0)\dagger} (y^\pm) \vert 0_F (\beta,\mu ) \rangle\,=\,
e^{\,-\,i\,\mu\,( x^\pm\, -\, y^\pm )} \langle 0_F (\beta ) \vert \psi^{(0)} (x^\pm ) \psi^{(0)\dagger} ( y^\pm) \vert  0_F (\beta ) \rangle\,.
\ee
Hence the dependence on $\mu$ factorizes in the form of phase factors. 
eq. (\ref{d2pf1}) is in agreement with the expression for the fermionic two-point function obtained in Ref. \cite{Min} following a different approach.

In the same way, one finds for the diagonal two-point function of the tilde fields

\be\label{d2pf2}
\langle 0_F (\beta,\mu ) \vert \widetilde\psi^{(0)} (x^\pm) \widetilde\psi^{(0)\dagger}(y^\pm) \vert 0_F (\beta,\mu ) \rangle
=\,e^{\,\,i\,\mu\,( x^\pm\, -\, y^\pm )} \langle 0_F (\beta ) \vert \widetilde\psi^{(0)} (x^\pm ) 
\widetilde\psi^{(0)\dagger} ( y^\pm) \vert  0_F (\beta ) \rangle\,,
\ee

For the off-diagonal thermal two-point function one has

\begin{eqnarray}\label{offdiag}
\langle 0_F (\beta,\mu ) \vert i\widetilde\psi^{(0)\dagger} (x^\pm) \psi^{(0)\dagger} (y^\pm) \vert 0_F (\beta,\mu ) \rangle\,&=&\,\langle 0 , \widetilde 0 \vert i \widetilde\psi^{(0)\dagger} (x^\pm ; \beta,\mu ) \psi^{(0)\dagger} ( y^\pm ; \beta,\mu ) \vert \widetilde 0 , 0 \rangle\,\nonumber\\
&=&
-\,\,\int_0^\infty\,\frac{dp}{2\pi}\,\left\{ e^{-\,i\,p\,(x^\pm\,-\,y^\pm )}\, \cos \theta (p; \beta,-\mu )\sin \theta (p ; \beta , -\mu )\,\right.\nonumber\\
&&\left.+\,e^{+\,i\,p\,(x^\pm\,-\,y^\pm)}\, \cos \theta ( p ; \beta , \mu ) \sin \theta (p; \beta,\mu )\,\right\}\,.
\end{eqnarray}

\no Using that 

\[
\sin \theta (p ; \beta , \pm\mu ) = e^{\,-\,\frac{\beta}{2}\,  (p \,\pm\,\mu )}\,\cos \theta (p ; \beta , \pm\mu )\,,
\]

\no one has

\begin{eqnarray*}\label{offdiag2}
\langle 0(\beta,\mu)\vert  i \widetilde\psi^{(0)\dagger} (x^\pm) \psi^{(0)\dagger}(y^\pm) \vert 0(\beta,\mu)\rangle\,&=&-\,\int_0^\infty\,\frac{dp}{2\pi}\,\left\{ e^{\,-\,i\,p\,(x^\pm - y^\pm)}\, \frac {e^{-\,\frac{\beta}{2} (p - \mu ) }}{1+e^{-\beta ( p-\mu)}}\,+\,e^{\,+\,i\,p\,(x^\pm-y^\pm) }\, \frac {e^{\,\frac{\beta}{2} (p + \mu )}}{1+e^{\beta(p+\mu)}}\,\right \}\,\\
&=& -\,\int_{-\infty}^\infty\,\frac{dp}{2\pi}\,\left\{ e^{\,-\,i\,p\,(x^\pm-y^\pm)}\, \frac {e^{\,-\,\frac{\beta}{2} ( p - \mu )}}{1+e^{-\beta(p-\mu)}}\,\right \}\,\\
&=&-\,e^{-\,i\mu ( x^\pm-y^\pm)}\,\int_{-\infty}^\infty\,\frac{dp}{2\pi}\,\left\{ e^{\,-\,i\,p\,(x^\pm\,-\,i\,\frac{\beta}{2}\,-y^\pm)}\, \frac 1{1+e^{-\beta p}}\,\right\}\,,
\end{eqnarray*}

\no which means that

\be\label{od2pf}
 \langle 0_F (\beta,\mu ) \vert i\,\widetilde\psi^{(0)\dagger} (x^\pm) \psi^{(0)\dagger} (y^\pm) 
\vert 0_F (\beta,\mu)\rangle\,=\,
e^{\,-\,i\,\mu\,( x^\pm\, -\, y^\pm )}\,
\langle 0_F (\beta ) \vert i\,\widetilde\psi^{(0)\dagger} (x^\pm ) \psi^{(0)\dagger} ( y^\pm) \vert  0_F (\beta ) \rangle\,,
\ee

\no where the thermal off-diagonal two-point function in the absence of the chemical potential is given by \cite{ABRI}

\be\label{analitcont}
\langle 0_F (\beta ) \vert (-i) \widetilde\psi^{(0)\dagger} (x^\pm ) 
\psi^{(0)\dagger} ( y^\pm) \vert  0_F (\beta ) \rangle\,=\,\,\frac{\displaystyle \,1}{\displaystyle 2 i \beta\,\sinh \frac{\pi}{\beta}\,(x^\pm\,-\,i \frac{\beta}{2}\,-\,y^\pm\,+\,i\,\epsilon )}\,.
\ee

As stressed in Refs. \cite{ABRI,ABRII,ABRIII}, the shift in the argument of the off-diagonal two-point function (\ref{analitcont}) can be understood in the context of the real time formalism as the tilde field living on the lower branch of the ``time" integration contour localized at $Im (t)\,=\,-\,i\,\frac{\beta}{2}$.

For the generalization to the $2n$-point functions one finds 

$$
 \langle 0 (\beta,\mu ) \vert \psi^{(0)} (x^\pm_1) \cdots \psi^{(0)} (x^\pm_n)\,\psi^{(0)\dagger} (y^\pm_1) \cdots \psi^{(0)\dagger }(y^\pm_n) \vert 0 (\beta,\mu ) \rangle\,=\,
$$

\be\label{generalized-2npoint-a}
\exp{\left({\,\displaystyle -\,i\,\mu\,\big ( \sum_{i = 1}^n\,x^\pm_i\,-\,\sum_{j = 1}^n\,y^\pm_j \big )}\right)}\,\,\frac{\displaystyle \prod_{i \langle i^\prime }^n \Omega \big (x^\pm_i\,-\,x^\pm_{i^\prime} ; \beta \big )\,\prod_{j \langle j^\prime }^n \Omega \big ( y^\pm_j\,-\,y^\pm_{j^\prime} ; \beta \big )}{\displaystyle \prod_{i , j = 1}^n\,\Omega \big ( x_i^\pm\,-\,y_j^\pm\,-\,i\, \epsilon (x_i^0\,-\,y_j^0  )\, ; \beta \big )}\,,
\ee
and

$$
\langle 0 (\beta,\mu ) \vert i \widetilde\psi^{(0)\dagger} (x^\pm_1) \cdots i \widetilde\psi^{(0)\dagger} (x^\pm_n)\,\psi^{(0)\dagger}(y^\pm_1) \cdots \psi^{(0)\dagger} (y^\pm_n) \vert 0 (\beta , \mu) \rangle\,=\,
$$

\be\label{generalized-2npoint-b}
(\,-\,1)^n\,\exp{\left(\,\displaystyle +\,i\,\mu\,\big ( \sum_{i = 1}^n\,x^\pm_i\,-\,\sum_{j = 1}^n\,y^\pm_j \big )\right)}\,\,\frac{\displaystyle \prod_{i < i^\prime }^n \Omega \big ( x^\pm_i\,-\,x^\pm_{i^\prime}\,-\,i\,\frac{\beta}{2} ; \beta \big )\,\prod_{j < j^\prime }^n \Omega \big ( y^\pm_j\,-\,y^\pm_{j^\prime}\,-\,i\,\frac{\beta}{2} ; \beta \big )}{\displaystyle \prod_{i , j = 1}^n\,\Omega \big ( x_i^\pm\,-\,y_j^\pm\,-\,i\,\frac{\beta}{2}\,-\,i\, \epsilon (x_i^0\,-\,y_j^0)\, ; \beta \big )}\,,
\ee

\no where

\[
\Omega (x ; \beta)\,=\,2\, i\, \beta\,\sinh \big ( \frac{\pi}{\beta}\,x\, \big )\,.
\]

Although the Lorentz invariance is not manifest in a field theory at finite temperature and we are not dealing with genuine Wightman functions, but with statistical ensemble thermal averages, one can follow the philosophy of the reconstruction theorem \cite{SW} as a heuristic guide line to reconstruct from the two-point function  the thermal Fermi field operator. From (\ref{generalized-2npoint-a}) and (\ref{generalized-2npoint-b}) one concludes that in the presence of temperature and chemical potential the massless chiral Fermi thermofields effectively factorize as follows:

\be\label{mu1-factorization}
{\cal U}_F^{-1}[\theta_F(\beta,\mu)]\psi^{(0)}(x^\pm){\cal U}_F[(\theta_F(\beta,\mu)]=
\psi^{(0)}(x^\pm;\beta,\mu) = e^{-i\mu x^\pm}\psi^{(0)}(x^\pm;\beta),
\ee 
\be\label{mu2-factorization}
{\cal U}_F^{-1}[\theta_F(\beta,\mu)]\widetilde\psi^{(0)}(x^\pm){\cal U}_F[(\theta_F(\beta,\mu)]
=\widetilde\psi^{(0)}(x^\pm;\beta,\mu) = e^{+ i\mu x^\pm}\widetilde\psi^{(0)}(x^\pm;\beta),
\ee
still obey the equations of motion of free massless fields since they are left or right moving:

\be\label{eqf1}
 i \gamma^\nu \partial_\nu \,  \psi^{(0)} (x ; \beta , \mu )\,=\, i \gamma^\nu \partial_\nu \,  
\widetilde\psi^{(0)} (x ; \beta , \mu )\,=\,0\,.
\ee

\subsection{ KMS conditions} 

Let $A_K(t)$ and $B_K(t)$ be operators evolving with time according to the ``Hamiltonian" $K=H_0 -\mu Q$
in (\ref{Boltzmann-average}):
\be\label{K-timeevolusion}
A_K(t) = e^{iKt}A(0)e^{-iKt}
\ee
and similarly for $B_K(t)$. Then, using the cyclical property  of the trace, we have
\be
Tr \left(e^{-\beta K}A_K(t)B_K(t')\right)=
Tr \left(e^{-\beta K}B_K(t')A_K(t+i\beta)\right)\,.
\ee
or in form of the Boltzmann average (\ref{Boltzmann-average}),
\be\label{KMS0}
\langle A_K(t)B_K(t^\prime)\rangle=\langle B_K(t^\prime)A_K(t+i\beta)\rangle.
\ee
This is the Kugo-Martin-Schwinger condition \cite{KMS}. Let us now translate this relation to
the operators evolving with the Hamiltonian $H$. With $[Q,H]=0$ we have the following relation between the two pictures:
\be\label{K}
A_K(t)=e^{- i \mu {\cal Q} t}A_H (t)e^{ i \mu {\cal Q}_t}=e^{-i\mu Q_A t}A_H(t)\,,
\ee
where ${\cal Q}_A$ is the charge associated with the operator $A$ as defined by $[{\cal Q},A]={\cal Q}_A A$. In terms of the operators $A_H (t) $ the KMS condition reads
\be\label{KMS1}
\langle A_H(t)B_H(t^\prime)\rangle=e^{\mu\beta Q_A}\langle B_H(t^\prime)A_H(t+i\beta)\rangle.
\ee
\no The fields $\psi^{(0)}(x^\pm)$ evolve in time with $H_0$. From 
(\ref{d2pf1}) and (\ref{psi-correlator})  one verifies that

\be\label{KMS2}
\langle 0(\beta,\mu)\vert\psi^{(0)}(x^\pm)\psi^{(0)\dagger} (y^\pm)\vert 0(\beta,\mu)\rangle=
e^{-\mu\beta}    \langle 0(\beta,\mu)\vert\psi^{(0)\dagger}(y^\pm)\psi^{(0)} (x^\pm +i\beta)\vert 0(\beta,\mu)\rangle\,,
\ee
which agrees with (\ref{KMS1}) since $[{\cal Q},\psi^{(0)}(x^\pm)]=-\psi^{(0)}(x^\pm)$.

Let us call $\psi^{\prime(0)}$ the two-component field evolving according to $K$. From (\ref{mu1-factorization}) we have

\be\label{psiprime}
\psi^{\prime(0)}(x;\beta,\mu)=e^{i\mu x^0}\psi^{(0)}(x;\beta,\mu)=
\left(
\begin{array}{c}
e^{-i\mu x^1}\psi^{(0)}(x^+;\beta)\cr
e^{+i\mu x^1}\psi^{(0)}(x^-;\beta)
\end{array}
\right)
\label{psi(0)-prime}
\ee
For these operators relation (\ref{KMS2}) reads

\be\label{KMS3}
\langle 0,\widetilde 0\vert\psi^{\prime (0)}_\alpha (x;\beta ,  \mu ) \psi^{\prime (0) \dagger}_\beta (y;\beta,\mu)\vert 0,\widetilde 0\rangle =
\langle 0,\widetilde 0\vert \psi^{\prime (0)\dagger}_\beta(y; \beta , \mu )\psi^{\prime (0)}_\alpha (x^0+i \beta , x^1 ; \beta , \mu )\vert 0,\widetilde 0\rangle
\ee
in agreement with condition (\ref{KMS0}).
We observe that $\psi^{\prime(0)}$ satisfies the modified equations of motion

\be\label{eqm1}
\left( i \gamma^\nu \partial_\nu +\mu\gamma^0\right)\,  \psi^{\prime(0)} (x ; \beta , \mu )\,=\,\left(- i \gamma^\nu \partial_\nu \, +\mu\gamma^0\right)\, \widetilde\psi^{\prime(0)} (x ; \beta , \mu )\,=\,0\,,
\ee

\no  in alignment with the prescription given in ref. \cite{Actor}.

\subsection{Bosonized formulation}
\bigskip\noindent

From the factorization property (\ref{mu1-factorization}) and (\ref{mu2-factorization}) we can infer the bosonized form of the free, zero mass fermion field at non-zero temperature and chemical potential. Indeed, as shown in \cite{ABRI,ABRII}, the zero mass free fermion field at finite temperature and vanishing chemical potential has the boson representation in terms of the Wick ordered exponential
(see ref \cite{AAR}; we omit the multiplicative phase factor which plays here no role.)
\be\label{mu-zero-psi}
\psi^{(0)}(x^\pm;\beta) = \sqrt{\left(\frac{\lambda}{2\pi}\right)}\dt e^{i\sqrt\pi\phi(x^\pm;\beta)} \dt\hskip 1cm \mbox{and} \hskip 1cm 
\widetilde\psi^{(0)}(x^\pm;\beta) = \sqrt{\left(\frac{\lambda}{2\pi}\right)}\dt e^{i\sqrt\pi\widetilde\phi(x^\pm;\beta)} \dt\,,
\ee
where $\lambda$ is an infrared regulator. Here $\phi(x^\pm;\beta)$ are the right- and left-moving components of a zero mass scalar field at inverse temperature $\beta$,

 \[
\phi(x^\pm;\beta) = {\cal U}^{-1}_B[\theta_B(\beta)]\phi(x^\pm){\cal U}_B[\theta_B(\beta)]
\]

\no where ${\cal U}_B[\theta_B(\beta)]$ is the unitary operator 
\[
{\cal U} [\theta_B (\beta ) ]\,=\,e^{\,-\,\int_{- \infty}^{+ \infty}\,dp^1\,\big ( \tilde a (p^1)\,a (p^1)\,-\,a^\dagger (p^1)\,\tilde a^\dagger (p^1) \big )\,\theta_B ( \vert p^1 \vert , \beta )}\,,
\]

\no with the Bogoliubov parameter $\theta_B ( \vert p^1 \vert ; \beta )$  implicitly defined by

\[
\sinh \theta_B (\vert p^1 \vert , \beta )\,=\,
\frac{ e^{\,-\, \frac{1}{2}\beta\,\vert p^1 \vert }}
 { \sqrt{1\,-\,e^{\, -\,\beta\,\vert p^1 \vert }}}\,,
\]

\[
\cosh \theta_B (\vert p^1 \vert , \beta )\,=\,
\frac{1}
 { \sqrt{1\,-\,e^{\, -\,\beta\,\vert p^1 \vert }}}\,.
\]

\no In particular we have for the 2-point function of $\phi(x^\pm;\beta)$ \cite{ABRII},
\[
\langle 0 , \widetilde 0 \vert \phi (x^\pm ; \beta ) \phi (0 ; \beta ) \vert \widetilde 0 , 0 \rangle \,=\,D^{(+)} (x^\pm ; \beta ,\lambda )\,=\,-\,\frac{1}{4 \pi}\,\ln \Big \{i \lambda \frac{\beta}{\pi}\,\sinh \frac{\pi}{\beta}(x^\pm - i \epsilon ) \Big \}\,+\,\frac{1}{2\pi} {\it z} (\beta \lambda)\,,
\]
where
\footnote{For sake of economy of parameters we choose the parameter $\mu^\prime = \mu$ (alias $\lambda$) in
\cite{ABRII}.}
\[
{\it z}(\beta\lambda)=\int_{\lambda}^\infty \frac{dp}{p}\frac{1}{e^{\beta p} - 1}.
\]
At finite chemical potential $\mu$ we thus have from (\ref{psiprime}) for $\psi'(x^\pm;\beta,\mu)$ the representation

\be\label{psiprime-bosonic}
\psi^{\prime(0)}( x;\beta,\mu) = \sqrt{\left(\frac{\lambda}{2\pi}\right)}\pmatrix{
e^{- i\mu x^1}:e^{i\sqrt\pi \phi (x^+;\beta)}:\cr
e^{+ i\mu x^1}:e^{i\sqrt\pi \phi (x^-;\beta)}:}\,,
\ee

\no and for the bosonic thermal vacuum

\[
\vert 0(\beta)\rangle = {\cal U}_B[\theta_B(\beta)\vert 0,\tilde 0\rangle.
\]
Note that this vacuum does not depend on the chemical potential. This is not surprising since the bosonic excitations are neutral. Note also that the transition from $\mu=0$ to $\mu\neq 0$ can be seen as a shift
$\phi(x^\pm) \to \phi(x^\pm)  - \frac\mu{\sqrt\pi} x^\pm$ formally generated by the time independent unitary operator
\footnote{One has for the left and right-moving component of a zero mass pseudoscalar field
$\phi(x) = \phi(x^+) + \Phi(x^-)$,
\[
[\phi(x^\pm),\partial_\pm \phi(y^\pm)] = i\delta(x^\pm - y^\pm)\,.
\]}
\[
U[\mu]=e^{-i \frac\mu{\sqrt\pi} \int_{-\infty}^{\infty}\left( y^+ \partial_+\phi(y^+)dy^+ +  y^-\partial_-\phi(y^-)dy^-\right)} \,.
\]

\subsection{Chirality dependent densities}

Let us present here a brief summary of the generalization of the  above results  to the case of independent chemical potentials for each chiral component. Instead of the Boltzmann average (\ref{Boltzmann-average}) we consider now

\be\label{Boltzmann-average1}
\langle A \rangle   = \frac{Tr  Ae^{-\beta(H_0 - \mu . {\cal Q})}}{Tr e^{-\beta(H_0 - \mu . {\cal Q})}},
\ee
where $\mu . {\cal Q}=\mu {\cal Q} +\mu_5 {\cal Q}_5$, with

\[
{\cal Q}_5 = \int_{-\infty}^\infty {dp^1}[b^\dagger(p^1)b(p^1) - d^\dagger(p^1)d(p^1)]\epsilon(p^1).
\]

Let us define  $\mu_\pm=\mu\mp\mu_5$. The thermalized operators are obtained as in eq. (2.2) with a change
in the chemical potential according to the signal of $p^1$.  All we have to do is to replace $\mu$ by
$\mu_-\Theta(p^1) + \mu_+\Theta(-p^1)$. When translated to the fermion components this amounts the change $\mu\longrightarrow \mu_\pm$ for
$\psi(x^\pm)$, respectively. It is straightforward to derive the factorization property for the fermionic operators in this case:

\be
U_F^{-1}[\theta_F(\beta,\mu_+,\mu_-)]\psi^{(0)}(x^\pm)U_F[(\theta_F(\beta,\mu_+,\mu_-)]=
\psi^{(0)}(x^\pm;\beta,\mu_\pm) = e^{-i\mu_\pm x^\pm}\psi^{(0)}(x^\pm;\beta),
\ee 
\be
U_F^{-1}[\theta_F(\beta,\mu_+,\mu_-)]\widetilde\psi^{(0)}(x^\pm)U_F[(\theta_F(\beta,\mu_+,\mu_-)]=
\widetilde\psi^{(0)}(x^\pm;\beta,\mu_\pm) = e^{+ i\mu_\pm x^\pm}\widetilde\psi^{(0)}(x^\pm;\beta).
\ee

The  operators satisfying the alternative KMS conditions evolving  according to $K\longrightarrow H_0-\mu Q-\mu_5 Q_5$,  are now given by

\be
\psi^{\prime(0)}(x;\beta,\mu_+,\mu_-)=e^{i\mu x^0+i\mu_5\gamma^5 x^0}\psi^{(0)}(x;\beta,\mu_+,\mu_-)=
\left(
\begin{array}{c}
e^{-i\mu_+ x^1}\psi^{(0)}(x^+;\beta)\cr
e^{+i\mu_- x^1}\psi^{(0)}(x^-;\beta)
\end{array}
\right).
\ee
The corresponding bosonized expression become

\be
\psi^{\prime(0)}( x;\beta,\mu_+,\mu_-) = \sqrt{\left(\frac{\lambda}{2\pi}\right)}\pmatrix{
e^{- i\mu_+ x^1}:e^{i\sqrt\pi \phi (x^+;\beta)}:\cr
e^{+ i\mu_- x^1}:e^{i\sqrt\pi \phi (x^-;\beta)}:}.
\ee
Note that the shift $\phi(x^\pm) \to \phi(x^\pm)  - \frac{\mu_\pm}{\sqrt\pi} x^\pm$ leads from $\psi^{(0)}(x^\pm;\beta)$ to $\psi^{(0)}(x^\pm;\beta,\mu_\pm)$.
\section{$QED_2$ at finite temperature and chemical potential}
The Dirac and Maxwell equations for $T=0$ read
\[
i \gamma^\mu \partial_\mu \psi (x)\,+\,\frac{e}{2}\, \gamma^\mu\,\lim_{{\varepsilon \rightarrow 0}\atop{\varepsilon^2 \langle 0}} \Big \{\,{\cal A}_\mu (x + \varepsilon ) \psi (x)\,+\,\psi (x)\,{\cal A}_\mu (x - \varepsilon )\,\Big \} = 0\,,
\] 
and 
 \[
 \partial_\mu { F}^{\mu\nu}+e{\cal J}^\nu=0
 \]
respectively.  The above operators are in the representation $A_H(x)$ in (\ref{KMS1}). For non-zero temperature but vanishing chemical potential, the solution to these equations in the Lorentz gauge has been shown to be given by  \cite{ABRIII}
\be\label{QED-fermion}
\psi (x;\beta) = \dt e^{\textstyle\,i \sqrt \pi \gamma^5 [ \Sigma (x;\beta) + \eta (x;\beta)]} \dt \psi^{(0)} (x;\beta)\,,
\ee
\be\label{QED-gauge-field}
{\cal A}_\mu (x;\beta)\, =\, -\, \frac{\sqrt \pi}{e} \epsilon_{\mu \nu} \partial^\nu 
\Big ( \Sigma (x;\beta) + \eta (x;\beta) \Big )\,,
\ee
where $\Sigma(x;\beta)$ is a massive free field of mass $m=\frac{e}{\sqrt\pi}$,
\[
(\Box+m^2) \Sigma(x;\beta)=0
\]
with two-point function
\footnote{The Bogoliubov parameter $\vartheta(p;\beta) $ is intrinsically defined by
\[
\cosh\vartheta(p;\beta) = \frac{1}{\sqrt{1-e^{-\beta p^0}}}\,,\quad
p^0 = \sqrt{p^2 + m^2}.
\]
}
(see the appendix of \cite{ABRIII})
\be\label{2-pt-Sigma}
\langle 0 , \widetilde 0 \vert \Sigma (x ; \beta ) \Sigma (y ; \beta ) \vert \widetilde 0 , 0 \rangle\,
=\,\Delta^{(+)}(x-y;\beta,m)
\ee
\[
= \,\frac{1}{4 \pi}\,\int_0^\infty \frac{d p}{\sqrt{p^2 + m^2}}\Big \{ e^{\,-\,i\,p^\mu (x - y)_\mu}\,\cosh^2 \vartheta (p ; \beta )\,+\,e^{\,i\,p^\mu (x - y)_\mu}\,\sinh^2 \vartheta (p ;\beta )\Big \}
\]
\be\label{periodic-sigma}
= \frac{1}{2 \pi}\,\sum_{n = -\infty}^\infty\,\int_0^\infty \frac{dp}{\sqrt{p^2 + m^2}}\,\cos (p x^1 )\,\Big (\,e^{\,-\, [\,n  \beta\,+\,i\,x^0 ] \sqrt{p^2 + m^2}}\,\Big )\,.
\ee
Here
$\psi^{(0)}(x;\beta)$ is the zero mass free fermion field (\ref{mu-zero-psi}) discussed in the previous section, and $\eta(x;\beta)$ is a zero mass field of indefinite metric with the two-point function \cite{ABRII}
\[
\langle 0 , \widetilde 0 \vert \eta (x^\pm ; \beta ) \eta (0 ; \beta ) \vert \widetilde 0 , 0 \rangle \,=\,+\,\frac{1}{4 \pi}\,\ln \Big \{i \bar\lambda \frac{\beta}{\pi}\,\sinh \frac{\pi}{\beta}(x^\pm - i \epsilon ) \Big \}\,-\,\frac{1}{2\pi} {\it z} (\beta \bar\lambda)\,,
\]
where $\bar\lambda$  is a small, but non-vanishing infrared regulator.

We now introduce the chemical potential, using as a guide that the corresponding two point function should satisfy the KMS condition in the form
of eq. (\ref{KMS2}). Since the fermion number ($Q$ in eq.(\ref{Boltzmann-average})) is only carried by the free massless fermion, we replace the free fermion in (\ref{QED-fermion}) by
\footnote{This heuristic argument can also be easily shown to follow from (\ref{Boltzmann-average}) with $H_0$ replaced by
$H=H_0 + \int_{-\infty}^{\infty} dp^1 \sqrt{(p_1^2 + m^2)}a^\dagger(p^1)a(p^1)$ and $Q$ the bare charge operator (\ref{charge-operator}).}
(see (\ref{psiprime}))
\[
\psi^{\prime (0)}(x;\beta,\mu)= e^{i\mu \gamma^5 x^1}\psi^{(0)}(x;\beta).
\]
This implies the change  $\psi(x;\beta)\longrightarrow\psi^\prime(x;\beta,\mu)$:
\be\label{solution1}
\psi^\prime (x;\beta,\mu) = \dt e^{\textstyle\,i \sqrt \pi \gamma^5 [ \Sigma (x;\beta) + \eta (x;\beta)]} \dt e^{i\mu \gamma_5 x^1}\psi^{(0)} (x;\beta)\,,
\ee
The corresponding fermion two-point function then reads
\be\label{QED-fermion2pointfunction}
\langle 0,\widetilde 0\vert \psi^\prime(x;\beta,\mu){\psi^\prime}^\dagger(y;\beta,\mu)\vert\widetilde 0,0\rangle =\langle 0,\widetilde 0\vert \psi^{\prime(0)}(x;\beta,\mu){\psi^{\prime(0)}}^\dagger(y;\beta,\mu)\vert\widetilde 0,0\rangle \times e^{\pi\gamma^5_x\gamma^5_y\left(\Delta^{(+)}(x-y;\beta,m)-D^{(+)}(x-y;\beta,\bar\lambda)\right)}
\ee
We next observe that the two-point function (\ref{2-pt-Sigma}) of  $\Sigma(x;\beta)$ is strictly periodic in imaginary time with period $\beta$, as seen from (\ref{periodic-sigma}).
The same property is inherited by the massless fields $\phi$ and $\eta$. Since
$\psi^{\prime(0)}$ satisfies the KMS condition  (\ref{KMS0}), the same applies thus to the full thermofield solution of $QED_2$.
Note also that $\psi^{\prime}(x;\beta,\mu)$ satisfies the modified Dirac equation
\be\label{Diraceq-K}
i \gamma^\mu \partial_\mu \psi^\prime (x;\beta,\mu)\,+\,\frac{e}{2}\, \gamma^\mu\,\lim_{{\varepsilon \rightarrow 0}\atop{\varepsilon^2 \langle 0}} \Big \{\,{\cal A}_\mu (x + \varepsilon;\beta ) \psi^\prime (x;\beta,\mu)\,+\,\psi^\prime (x;\beta,\mu)\,{\cal A}_\mu (x - \varepsilon;\beta )\,\Big \}+\mu\gamma^1\gamma^5\psi^\prime(x;\beta,\mu) = 0\,.
\ee
The additional term proportional to $\mu$ reflects the fact that the time development of the fermion operator is now determined by the operator $H-\mu {\cal Q}$. 
Remembering that $\gamma^1\gamma^5=\gamma^0$, we see that this equation is in agreement with the prescription given in \cite{Actor}, which  has been the starting point in  \cite{Kogut, Kao,Schaposnik}.  If we were to  use $\psi^{(0)}$ instead of $\psi^{\prime(0)}$, we would obtain a thermofield evolving according to $H$ 
satisfying the KMS condition in the form of (\ref{KMS2}).

It remains to examine the ``Maxwell equation" satisfied by the gauge field. 
The  current associated to $\psi^\prime(x;\beta,\mu)$ can be computed as the symmetrized, space-like  short distance limit
(we suppress the $\beta$ and $\mu$ arguments) 
\[
{\cal J}^\prime_\nu (x)=lim_{\varepsilon\to 0}Z^{-1}(\varepsilon)\frac12 \left(\bar\psi^\prime(x+\varepsilon)\gamma_\nu e^{ie\int_x^{x+\varepsilon}dz^\mu A_\mu }\psi^\prime(x)+\bar\psi^\prime(x)\gamma_\nu e^{ie\int_{x-\varepsilon}^x dz^\mu A_\mu }\psi^\prime(x-\varepsilon)\right)
\]
Using the representation given in eq. (\ref{solution1}) one finds after a multiplicative renormalization (for an analogous sample calculation the reader should consult \cite{AAR}, section
10.2.2)
\be\label{current}
{\cal J}^\prime_\nu (x)=\frac{-1}{\sqrt\pi}\epsilon_{\nu\lambda}\partial^\lambda\left(\Sigma + \phi+\eta\right)+\frac\mu\pi\delta_{\nu 0}\,,
\ee
so that $<{\cal J}^\prime_\nu> = \frac{\mu}{\pi}\delta_{\nu 0}$\,.
On the other hand we have from (\ref{QED-gauge-field}),
\[
\partial_\mu F^{\mu\nu}=\frac{-1}{\sqrt\pi}\epsilon_{\mu\nu}\partial^\nu\Sigma,
\]
so that the Maxwell equation will be satisfied in a weak form on the subspace $|\Psi\rangle$  defined by $\partial_\mu(\eta+\phi)|\Psi\rangle=0$, provided we add  a term
$\mu A^0$, representing a uniform density of background charges, to the Lagrangian, as has been done in \cite{Kao}. Notice that with the usual boundary conditions at $t=\pm\infty$, this term does not affect
the gauge invariance of the corresponding action.

The consideration of independent chemical potential for each chiral components would impact in a term added to the left hand side of eq.
(\ref{current}) leading to $<J_\nu> = \frac{\mu}{\pi}\delta_{\nu 0}+\frac{\mu_5}{\pi}\delta_{\nu 1}$. This leads to a persistent current, a phenomenon which  has been noticed in \cite{Min}.

\section{Correlators of chiral condensates in the theta vacuum} 

In \cite{Schaposnik} the correlators of chiral densities in $QED_2$ have been computed at zero temperature but finite chemical potential, using euclidean functional
integration methods, taking account of the non-trivial topological field configurations reflecting  the existence of an infinitely
degenerate  vacuum, and allowing for the tunnelling between different topological vacua \cite{RS}. In this section we shall address this problem at finite temperature and chemical potential from an operator point of view  within Thermofield Dynamics. As will be seen, the results can be presented in
a compact way if one works in the so called theta vacuum.

The details behind the following calculations are contained in  ref \cite{ABRIII}. The reader is invited to consult this reference  as well as chapters X and XII of refs.\cite{AAR}.

The chiral densities $(\bar{\psi}^\prime_{\alpha} (x;\beta,\mu)\psi^\prime_\alpha(x;\beta,\mu))$ ($\alpha = 1,2$
) are gauge invariant. 
For their calculation it is useful to make a convenient choice of gauge. The gauge transformations are generated by the longitudinal part of the, $\mu=0$, current,  \cite{RS1}.
 We shall  choose the so called ``$\sqrt\pi$" gauge \cite{Lowenstein} where 
\[
{\cal A}_\mu (x;\beta)\, =\, -\, \frac{\sqrt \pi}{e} \epsilon_{\mu \nu} \partial^\nu  \Sigma (x;\beta)\,.
\]   
In this gauge we shall denote the operator $\psi^\prime$ by $\Psi^\prime$. We have
\[
\Psi^\prime(x;\mu)=(\frac{\lambda}{2\pi})^{\frac12}: e^{i\sqrt\pi\gamma^5\Sigma (x)}: e^{i\mu\gamma^5 x^1}\sigma,
\]
\noindent where $\sigma_\alpha (\alpha=1,2)$ are constant ``spurious"  $T=0$ operators merely carrying the the conserved charge and chirality quantum numbers. These ``spurionic constant"  operators commute
among themselves as well with all observables, and their repeated application on the $\vert 0,\widetilde 0\rangle$ vacuum 
leads to an infinite number of degenerate thermal vacua \cite{ABRIII}.  The``$\sqrt\pi$" gauge thus explicitely displays the
 massive Sigma field as the only physical degree of freedom in an infinite "sea" of vacua.

Consider now the chiral  densities 
\[
S_\pm=:\bar\Psi^\prime \frac{1\pm \gamma^5}2\Psi^\prime:.
\]
We have 
\footnote{$\sigma_\alpha$ without explicit argument $\beta$ refers to temperature zero.}
\[
S_\pm(x)=(\frac{\lambda}{2\pi}): e^{\pm i2\sqrt\pi\Sigma(x)}: e^{\pm 2i\mu x^1}\sigma_\pm
\]
\noindent with $\sigma_+=\sigma_1^\dagger\sigma_2$ and $\sigma_-=\sigma_+^\dagger$.

For the calculation of the correlators of these densities we consider the following coherent superposition of vacuum states:
\footnote{Since the total Hamiltonian $H-\hat H$ commutes with $U[\beta]$, this is a possible choice of zero energy eigenstates of the total Hamiltonian, particularly convenient for our calculations.}
\[
|\theta;\beta\rangle=\sum_{n=-\infty}^{\infty} e^{-i n\theta}\sigma_+^n |0(\beta)\rangle.
\]
\noindent This represents only a subspace of the full set of vacuum states, as can be seen from from the discussion in \cite{ABRIII}.
Noting that
\[
\sigma_+\vert\theta;\beta \rangle= e^{ i\theta}|\theta;\beta\rangle\,,\quad
\langle \theta,\beta\vert\sigma_- = \langle \theta,\beta\vert  e^{-i\theta}
\]
we have
\[
{\langle\bar \Psi^\prime \Psi^\prime\rangle}_\theta=\langle S_++S_-\rangle_\theta=2\frac \lambda{2\pi}\cos{(2\mu x^1+\theta)}
\]
Using further
\[
S_\pm(x_1)S_\pm(x_2)=\left(\frac\lambda{2\pi}\right)^2 \sigma_\pm^2 e^{\pm 2i\mu(x^1_1+x^1_2)}e^{-4\pi \Delta^{(+)}(x_1-x_2);\beta,m}:e^{\pm i2\sqrt{\pi}\sum_{i=1}^2\Sigma(x_i)}:
\]
and
\[
S_\pm(x_1)S_\mp(x_2)=\left(\frac\lambda{2\pi}\right)^2 \sigma_+\sigma_- e^{\pm 2i\mu(x^1_1-x^1_2)}e^{4\pi \Delta^{(+)}(x_1-x_2)}
:e^{\pm i2\sqrt{\pi}(\Sigma(x_1)-\Sigma(x_2))}:\,,
\]
one finds for the expectation values in the thermal theta vacuum $|\theta,\beta>$,
\[
\langle(\bar\Psi^\prime(x_1)\Psi^\prime(x_1))(\bar\Psi^\prime(x_2)\Psi^\prime(x_2))\rangle_\theta=
\left(\frac\lambda{2\pi}\right)^2  2\left(\cos{( 2\mu(x^1_1+x^1_2)+2\theta}\right)e^{-4\pi \Delta^{(+)}(x_1-x_2;\beta,m)}
\]
\[
\left.+\cos{\left( 2\mu(x^1_1-x^1_2)\right)}e^{4\pi \Delta^{(+)}(x_1-x_2;\beta,m)}\right.
\]
Let us note that the chemical potential brakes the invariance under translation due to the factors
$\exp{i\mu x^1}$ appended to the fields. The solution thus depends on the choice of the origin. This dependence  can be associated to a choice of
the origin of the charge distribution in space. It is interesting to note that the $\theta$ parameter emerging in the above formulae
can be viewed as a redefinition of the origin of this charge distribution.

The above result may be easily generalized. It is not hard to see that for the correlator of $n$ chiral densities we have
\[
\langle(\bar\Psi^\prime(x_1)\Psi^\prime(x_1))\cdots(\bar\Psi^\prime(x_n)\Psi^\prime(x_n))\rangle_\theta=\hspace{4cm}
\]
\[\hspace{2cm} 2\left(\frac{\lambda}{2\pi}\right)^n  
\sum_{e_i=\pm 1}\left\{\cos{\left(\sum_{i=1}^ne_i (  2\mu x^1_i+\theta)\right)}e^{-4\pi \sum_{i\langle j} e_ie_j \Delta^{(+)}(x_i-x_j;\beta)}\right\}\,.
\]

For the particular value $\theta=0$ (and zero temperature) this result agrees with that in ref. \cite{Schaposnik}. Let us note the striking simplicity of
this derivation in contrast to that using  functional integral methods.  Moreover, our results generalize those of  ref. \cite{Schaposnik} to arbitrary
theta vacua, as well as to finite temperature.

The consideration of independent chemical potentials, $\mu_\pm$, does not impact the computation of the chiral densities here discussed.

\section{Conclusion}
The discussion of $QED_2$ at finite temperature and chemical potential has made use in the past of bosonization correspondences \cite{C} and functional
techniques, combined with the prescription given in ref. \cite{Actor}. In the present case we have taken an operator approach based on Thermofield Dynamics, reminiscent of the operator techniques of Lowenstein and Swieca, but in a doubled Hilbert space. We have considered two forms of the solution, corresponding to two forms
of the KMS relation which the vacuum expectation values of two point functions should satisfy. The relation of these solutions to those discussed in the literature has been commented. In particular we have shown that the dependence on the chemical potential is entirely contained in the free fermion part. For this reason an entire section has been devoted to the discussion of the free fermion correlation functions in the presence of a chemical potential and to the construction of its bosonized representation. The n-point functions 
of chiral densities have been computed at finite temperature and chemical potential in a general $\theta$-vacuum, and have in particular been shown to agree with those  computed in ref. \cite{Schaposnik}
for zero temperature and $\theta=0$ vacuum. For $\theta \neq 0$, the theta dependence of the chiral condensates presents itself as a shift proportional to $\theta$ of the origin of the coordinate system. This restores the translational invariance of the expressions, absent in the case $\theta=0$.

 {\bf Acknowledgments}

{ The authors are grateful to the Brazilian Research Council (CNPq) and to the DAAD scientific exchange program, which made this collaboration possible. One of us (K.D.R.) wishes to thank the Department of  Physics of the Universidade Federal Fluminense (Brazil) for the kind hospitality extended to him.}



\begin{thebibliography}{16}
\bibitem{AAR} E. Abdalla, M.C.B. Abdalla and K.D.  Rothe,  "Non-Perturbative
Methods in two
dimensional Quantum Field Theory", World Scientific, Singapore, 1991 and 2001 (2nd edition).


\bibitem{ABRI} R. L. P. G. Amaral, L. V. Belvedere and K. D. Rothe, Ann. of Phys. {\bf 320} (2005) 399.



\bibitem{TFD} L. Leplae, F. Mancini and 
H. Umezawa, Phys. Rep. {\bf 10 C} (1974), 151; Y. Takahashi 
and H. Umezawa, {\it Collective Phenomena} {\bf 2} (1975, 55; H. Matsumoto, Fortschr.
Physik {\bf 25} (1977), 1.


\bibitem{U} H. Umezawa, H. Matsumoto and M. Tachiki, ``Thermo Field Dynamics
and Condensed States'', Nort-Holland, Amsterdam 1982.

\bibitem{Das} A. Das, ``Finite Temperature Field Theory", World Scientific
1997.

\bibitem{Mats1} H. Matsumoto, Y. Nakano and H. Umezawa, J. Math. Phys.
 {\bf 25} (1984), 3076.

\bibitem{O} I. Ojima, Ann. Phys. {\bf 137} (1981) 1; 

\bibitem{Mats} H. Matsumoto, I. Ojima and H. Umezawa, Ann. 
Phys. {\bf 152} (1984) 348.

\bibitem{ABRII} Amaral, R. L. P. G., Belvedere, L. V. and Rothe, K. D., Ann. of Phys. (2008)  2662.


\bibitem{C} S. Coleman, Phys Rev. {\bf D 11} (1975) 2088.



\bibitem{Haag} R. Haag, N. M. Hugenholtz and Winnink, Comm. Math. 
Phys. {\bf 5} (1967) 215.



\bibitem{KMS}
Kubo, R., J. Phys. Soc. Japan, {\bf 12} (1957) 570;
Martin, P. C. and Schwinger, J., Phys. Rev. {\bf 115} (1959) 1342;

\bibitem{Lowenstein}J. Lowenstein and J. A. Swieca,  Ann. of Phys. {\bf 68} (1971) 172.

\bibitem{ABRIII} L.V. Belvedere, R.L.P.G. Amaral, K.D. Rothe and A.F. Rodrigues, 
``Quantum Electrodynamics in Two-Dimensions at Finite Temperature; Thermofield Bosonization Approach", arXiv:0908.1558v1, to appear in J. Phys. A.

\bibitem{Min} A. Liguori, M. Mintchev and L. Pilo, Nucl. Phys. {\bf B 569}
(2000) 577.


\bibitem{Kogut}W. Fischler, J. Kogut and L. Susskind, Phys. Rev. {\bf D 19} (1979) 1188. 
\bibitem{Kao}Yeong-Chuan Kao and Yu-Wen Lee, Phys. Rev. {\bf D50} (1994) 1165.
\bibitem{Schaposnik} H. R. Christiansen,  F. A. Schaposnik, Phys. Rev. {\bf D 53} (1996) 3260.
\bibitem{Actor} A. Actor, Phys. Lett. {157B} (1985) 53.

\bibitem{RS}K. D. Rothe and J. A.  Swieca, Ann. Phys. 117 (1979) 382. 


\bibitem{SW} R. F. Streater and A. S. Wightman, ``PCT , SPIN ,$\&$ STATISTICS AND ALL THAT '' Mathematical Physics Monograph Series, W. A. Benjamin, INC, New York, 1964.
\bibitem{RS1} K.D. Rothe and J.A. Swieca, Phys. Rev. D15 (1977)541.


\end{thebibliography}
\end{document}